\documentclass[12pt,preprint]{aastex}

\usepackage{epsfig}

\shorttitle{Transients Event Rates}
\shortauthors{Macquart}

\begin{document}


\title{Detection Rates for Surveys for Fast Transients with Next Generation Radio Arrays}


\author{Jean-Pierre Macquart\altaffilmark{1}}
\affil{ICRAR/Curtin University, Bentley, WA 6845, Australia; J.Macquart@curtin.edu.au}


%


\altaffiltext{1}{ARC Centre of Excellence for All-Sky Astrophysics (CAASTRO)}

\begin{abstract}
We relate the underlying properties of a population of fast radio-emitting transient events to its expected detection rate in a survey of finite sensitivity.  The distribution of the distances of the detected events is determined in terms of the population luminosity distribution and survey parameters, for both extragalactic and Galactic populations.  The detection rate as a function of Galactic position is examined to identify regions that optimize survey efficiency in a survey whose field of view is limited.  The impact of temporal smearing caused by scattering in the Interstellar Medium has a large and direction-dependent bearing on the detection of impulsive signals, and we present a model for the effects of scattering on the detection rate.  We show the detection rate scales as $\Omega S_0^{-3/2 + \delta}$, where $\Omega$ is the field of view, and $S_0$ is the minimum detectable flux density, and $0 < \delta \le 3/2$ for a survey of Galactic transients in which interstellar scattering or the finite volume of the Galaxy is important.  We derive formal conditions on the optimal survey strategy to adopt under different circumstances for fast transients surveys on next generation large-element, widefield arrays, such as ASKAP, LOFAR, the MWA and the SKA, and show how interstellar scattering and the finite spatial extent of a Galactic population modify the choice of optimal strategy.   
\end{abstract}


\section{Introduction}

Short-timescale transients are often associated with the highest energy density events in the Universe. Known examples, such as pulsars, magnetars and Rotating RAdio Transients (RRATs), show that the emission from such impulsive events is usually generated by matter under extreme conditions whose properties probe physical regimes that far exceed those accessible in terrestrial laboratories.  Even without complete understanding of the radiation mechanism, the mere existence of such emission can be used to probe the behaviour of matter and spacetime under extreme conditions.  This is epitomized by the discovery of radio pulsars, and their subsequent use to test general relativity (Hulse \& Taylor 1974; Taylor \& Weisberg 1982) and the neutron star equation of state (e.g. Weber et al. 2009).

There is renewed impetus to detect new classes of short timescale radio emitting objects following the detection of a $\sim 30\,$Jy, 5-ms duration one-off pulse reported by Lorimer et al.\,(2007).  The frequency-time characteristics of the burst, if due to propagation through a dispersive medium, indicates a dispersion measure of 375\,pc\,cm$^{-3}$, which would place this putative object at a cosmological distance.  Its extraordinary luminosity and short duration generated a flurry of speculation as to its origin (e.g. Vachaspati 2008, Kavic et al. 2008), and the hope that more such objects would be useful as probes of the elusive low-redshift ionized inter-{\it galactic} medium, much in the same way that pulsars have proven exquisite probes of the interstellar medium of our Galaxy (Ginzburg 1973; Palmer 1993; Ioka 2003, Inoue 2004)
The astronomical provenance of the burst has recently been questioned (Burke-Spolaor et al.\,2011), but its origin will ultimately only be resolved once other examples of this burst are detected and followed up.     

Several large collaborations are conducting or planning surveys for fast radio transients.  This includes groups using LOFAR (Fender et al. 2006; Hessels et al. 2008), the VLBA (V-FASTR, Wayth et al. 2011; Thompson et al. 2011), Parkes (HTRU survey, Keith et al. 2010), ASKAP (CRAFT, Macquart et al. 2010), the MWA (2PiP, S. Ord, private communication) and MeerKat (TRAPUM, PIs Stappers \& Kramer).  Although highly sensitive in their own right, these surveys are envisaged as precursors to those that will ultimately be conducted with the SKA (e.g. Cordes et al. 2010).

A crucial element to the detection of rare events is the large field of view (FoV) afforded by the new technologies employed by these surveys; the Parkes telescope utilizes multibeam technology, the MWA and LOFAR use aperture array technology which can, in principle, detect objects over a large fraction of the visible sky, while APERTIF and ASKAP employ focal plane aperture array technology to achieve a FoV of 8 and 30\,deg$^2$ respectively.  In many of these telescopes both the spatial distribution of the interferometer elements and backend processing limitations force a tradeoff between sensitivity and the FoV that may be searched for transients.  

Unfortunately, little consideration has been devoted to the optimal tradeoff between sensitivity and FoV for arrays that seek to detect transient radio emission.  SKA Memo 100 (Schilizzi et al. 2007) explored the interaction between the FoV and sensitivity trade-offs for the cost of various SKA design concepts.  
Transients are an important and subtly different case because, as one-off events, one cannot trade integration time in the same way as for non-transient sources or pulsars.  This is particularly the case when the duration of a transient is much shorter than the dwell time of the telescope on each field.  This restriction particularly applies in the case of highly impulsive transients, whose total duration, including the dispersion sweep, is much shorter than the dwell time of the telescope on any individual pointing.  This typically refers to any object with duration shorter than $\sim 5$\,s.  


The goal in this paper is to relate the underlying properties of a population of transients to its expected detection rate and, for a Galactic population, its sky distribution after the geometry of the Galaxy and interstellar scattering are taken into account.  We also examine the distance distribution of detected objects within a survey; this consideration is particularly important for the detection of impulsive signals associated with fast transients, where a search over a range of dispersion measures is a necessary step in the detection process.  As this can represent a sizeable fraction of the computational cost of a survey, informed decisions must be made on the useful range of dispersion measures that should be searched.  The present analysis seeks to duplicate the spirit of the pulsar population synthesis simulations of Smits et al.\,(2009, 2011) which examine various pulsar survey strategies for the SKA, and in a similar vein, we wish to use the insight derived by our analysis to remark on optimal survey strategies for fast transients. 

The layout of this paper is as follows.  In \S\ref{sec:defns} we define the problem and place these calculations in the context of pre-existing performance metrics of transients surveys.  In  \S\ref{sec:extragalactic} we examine the event rate for a population of extragalactic transients, while in \S\ref{sec:galactic} we examine the corresponding statistics for a population of transients bound to our Galaxy.  We then compute how the effects of temporal smearing due to interstellar scattering, which is important at the operating wavelengths of forthcoming widefield radio arrays, alter the event detection rate for Galactic transients.  In \S\ref{sec:discussion} the implications of these results for forthcoming surveys for fast transients are discussed in terms of tradeoffs between survey FoV and sensitivity.  The conclusions are presented in \S\ref{sec:conc}.

\section{Definitions} \label{sec:defns}
Our objective is to compute the detection rate of transients, ${\cal R}$. 
We define the rate volume density of transient objects of some given category as $\rho_0$ with dimensions of events per unit time per unit volume.  The rate of transients observed over the full sky is, $\rho_0  \, V_{\rm max}$, where $V_{\rm max}$ is the limiting volume out to which the survey telescope could detect these objects, which depends on both the sensitivity of the telescope and on the luminosity of the objects under consideration.   The detection rate over a FoV $\Omega$ is then,
\begin{eqnarray}
{\cal R} = \rho_0 \, \frac{\Omega}{4 \pi} \, V_{\rm max} .
\end{eqnarray}

In the simple case of a homogeneous population of objects the maximum distance, $D_{\rm max}$, out to which an object of luminosity ${\cal L}_\nu$ could be detected is, 
\begin{eqnarray}
D_{\rm max} = \sqrt{\frac{{\cal L}_\nu}{4 \pi S_0}},
\end{eqnarray}
where,  
\begin{eqnarray}
S_{0} = \frac{ m k_B T_{\rm sys}}{A_e \sqrt{n_p \Delta \nu \, \Delta T} } , \label{eqS0}
\end{eqnarray}
is the telescope sensitivity on the duration of the transient outburst, $\Delta T$.  The sensitivity depends on the effective collecting area, $A_e$, the system temperature, $T_{\rm sys}$, the minimum detectable S/N, $m$, the number of polarizations recorded, $n_p$, and $k_B$ is Boltzmann's constant.  We take $S_0$ to be single-valued throughout this analysis, but note that if the transient events span a range of durations the effective sensitivity spans a range of values.  

The value of $S_0$ is impacted if the detection filter is not optimally matched to the timescale of the transient events.  We refer the reader to Cordes \& McLaughlin (2003) for a detailed discussion of the impact of non-optimal detection on performance.  Decreased performance associated with a detection filter that is not optimally matched to the burst shape and duration is easily accounted for in the present calculation by altering the threshold S/N for a detection, $m$ in eq.\,(\ref{eqS0}), as appropriate.  More generally, the effects of interstellar and (possibly) intergalactic scattering also degrade the sensitivity of a survey to bursts of sufficiently short duration.  Scattering acts to temporally smear out an impulsive signal, resulting in a distance-dependent degradation of the signal strength.  An explicit treatment of the effects of temporal smearing is deferred to a later section, where the analysis naturally takes into account the variation in $S_0$ associated with the fact that the apparent burst duration is larger\footnote{For the case of temporal smearing, it is assumed that the search software remains optimally matched to the burst duration. (I.e. that there is sufficient computational capacity for the search software to trail a range of detection windows of various durations, and that one such window closely matches the temporal duration of the temporally smeared burst.)}.

If the population is luminous enough to be detected at extragalactic distances and is homogeneously distributed one has $V_{\rm max} =  4 \pi D_{\rm max}^3 /3$.
However, the assumption that the population is homogeneously distributed across the Universe is violated for both for Galactic populations and for extragalactic populations in which there is significant evolution in the transients' properties with redshift.  If the population is primarily Galactic, $V_{\rm max}$ is ultimately bounded by the size of the Galaxy.

For objects spanning a large range of intrinsic luminosities it is useful to make the dependence on transients rate with luminosity explicit.  We define the rate volume and luminosity density of transients as $\rho_L$ so that the total rate of transients per unit volume is, 
\begin{eqnarray}
\rho_0 = \int_0^\infty \rho_{\cal L} \, d{\cal L}_\nu.
\end{eqnarray}

\subsection{The Source Luminosity Distribution} 

\begin{figure}[htb!]
\begin{center}
\epsfig{file=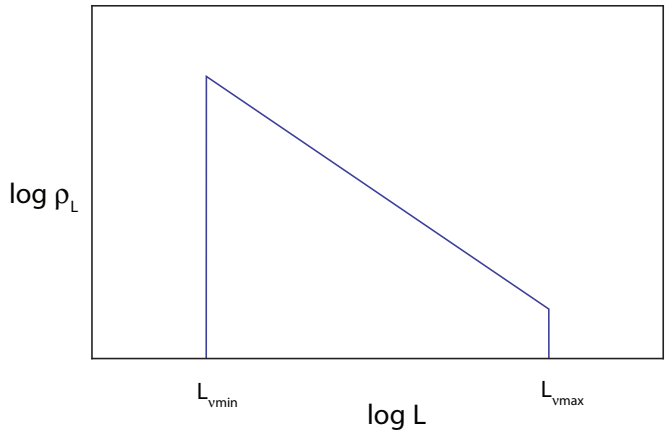,scale=0.6}
\caption{The generic luminosity distribution considered in the text.} \label{fig:LDistn}
\end{center}
\end{figure}

For many astrophysical objects the luminosity distribution follows a power-law with index $-\alpha$ between minimum and maximum luminosities, ${\cal L}_{\nu,{\rm min}}$ and ${\cal L}_{\nu,{\rm max}}$, respectively. The number of objects over a luminosity interval $d{\cal L}_\nu$ can be written as 
\begin{eqnarray}
\rho_{\cal L} d{\cal L}_\nu = 
\frac{\rho_0 \, d{\cal L}_\nu }{K} \,  \left\{
\begin{array}{l l} 
{\cal L}_{\nu}^{-\alpha}, & {\cal L}_{\nu,{\rm min}} < {\cal L}_{\nu} < {\cal L}_{\nu,{\rm max}},  \\ 
0, & \hbox{otherwise}.
\end{array}
\right. \label{Ldistn}
\end{eqnarray}
where $\rho_0$ is the total event rate per unit volume integrated over all luminosities and the normalization constant is, 
\begin{eqnarray}
K &=& \left\{ \begin{array}{ll}
 \frac{1}{\alpha -1} \left( {\cal L}_{\nu,{\rm min}}^{1-\alpha} - {\cal L}_{\nu,{\rm max}}^{1-\alpha} \right), & \alpha > 0 \,\, \& \,\, \alpha \neq 1, \\
 \ln \left( \frac{{\cal L}_{\nu,{\rm max}}}{{\cal L}_{\nu,{\rm min}}} \right), & \alpha =1. \\ 
\end{array} \right.
\end{eqnarray}
A schematic illustration of this luminosity distribution is shown in Fig.\,\ref{fig:LDistn}.   
The number of events detected in the luminosity range $[{\cal L}_\nu,{\cal L}_\nu + d{\cal L}_\nu]$ is then
\begin{eqnarray}
R_{\cal L} \, d{\cal L}_\nu = \rho_L \, \frac{\Omega}{4\pi} \, V_{\rm max}({\cal L}_\nu)\, d{\cal L}_\nu.
\end{eqnarray}
The volume of detectability, $V_{\rm max}$, is a function of luminosity, and the total event rate is found by integrating over all luminosities,
\begin{eqnarray}
{\cal R} &=& \int_{{\cal L}_{\nu,{\rm min}}}^{{\cal L}_{\nu,{\rm max}}} R_{\cal L} d{\cal L}_\nu \nonumber \\
&=& \frac{\Omega}{4\pi} \int_{{\cal L}_{\nu,{\rm min}}}^{{\cal L}_{\nu,{\rm max}}} \rho_{\cal L} V_{\rm max}({\cal L}_\nu) d{\cal L}_\nu.
\end{eqnarray}

Given that the sorts of objects a fast transients survey is sensitive to are almost certainly associated with nonthermal coherent emission, it is pertinent to place the luminosity function in the context of a known group of coherent radio emitters.  The exemplar of this class is pulsars.  Recent studies of the Galactic population suggest that, for pulsars, ${\cal L}_{\nu,{\rm min}} = 0.01\, $mJy\,kpc$^{2}$ and ${\cal L}_{\nu,{\rm max}} = 32$\,Jy\,kpc$^{2}$ with $\alpha = 1.2$--$2$ (Faucher-Gigu\`ere \& Kaspi 2006; Lorimer et al. 2006)\footnote{We note that, as these figures are based on surveys of a finite number of pulsars, it is quite plausible that a survey over a much larger volume (i.e. incorporating the pulsar luminosity distributions of other galaxies) would increase $L_{\nu,{\max}}$}.  The giant pulses from some well-studied individual pulsars are also exhibit power-law brightness distributions, with the  luminosity distributions of individual pulsars in the range $\alpha=1.8$ to $3.0$. 

\subsection{The distinction between detection rate and survey speed}
It is instructive to examine the difference between the detection rate and the oft-used survey Figure of Merit (FoM) as metrics of survey efficiency.  Consider by way of illustration the survey detection rate for a population of objects of fixed intrinsic luminosity (i.e. a population of standard candles) distributed homogeneously throughout space.  We shall consider two surveys possible with a telescope comprised of $N$ dish elements each of effective area $A_t$, each with FoV $\Omega_t$: (i) a ``collimated'' survey mode in which all telescopes are pointed at the same direction, and the total powers detected at each telescope are combined incoherently so that the total effective area pointed at each patch of sky is $N^{1/2} A_t$, (ii) a ``fly's-eye'' mode in which each telescope is pointed in a different direction so that the total FoV is $N \Omega_t$, but the total effective area for each pixel on the sky is only $A_t$. 
 
Following Cordes (2010), D'Addario (2010) defines the Figure of Merit for a transients survey as
\begin{eqnarray}
F_t &=& F_s \, {\cal K}(\eta W, \tau/W), \qquad \hbox {with}\\
F_s &=& \Delta \nu \, p \, \Omega_p \, \left( \frac{A_e}{T_{\rm sys}} \right)^2,
\end{eqnarray}
where $\Delta \nu$ is the bandwidth, $\Omega_p$ is the FoV per pixel, $p$ is the number of pixels, and the function ${\cal K}(\eta W,\tau/W)$ accounts for the fact that sources may be on for only a fraction of the dwell time, with the bursts being of duration $W$, and the burst event times follow a Poisson distribution with rate $\eta$. D'Addario (2010) argues that the value of ${\cal K}$ is identical for the purposes of comparing between cases of different telescope parameters, so that only the value of $F_s$, which is just the FoM for a survey of a population of steadily-emitting sources, need be considered.   

Compare $F_s$ between the two surveys.  In case (i) one has $F_s = \Delta \nu \Omega_t (N^{1/2} A_t/T_{\rm sys} )^2$, while in (ii)  one has $F_s = \Delta \nu N \Omega_t (A_t/T_{\rm sys})^{2}$.  Thus we conclude that, with $F_s$ the same, $F_t$, the survey FoM, is identical for the two cases and that there is no advantage to a fly's eye configuration over the collimated configuration of array elements.
It is unsurprising that the two survey configurations are indistinguishable in this instance since the FoM, $F_s$, only quantifies the rate of solid angle the telescope can survey down to a given sensitivity limit.  


Now consider the actual detection rate for the two survey strategies.  Suppose an individual antenna can detect an object down to a sensitivity ${\cal S}$.  The minimum detectable flux density from a configuration in which the total powers from the antennas are combined incoherently is $N^{-1/2} {\cal S}$.  For a population of transient sources distributed homogeneously throughout a volume, the detection rate for configuration (i) is 
\begin{eqnarray}
{\cal R}_i = \frac{4 \pi}{3} \rho \, \frac{\Omega_t}{4\pi} \ \left( N^{1/2} \frac{{\cal L}_\nu}{4 \pi {\cal S}} \right)^{3/2},
\end{eqnarray}
while the expected detection rate for the fly's eye, configuration (ii), is
\begin{eqnarray}
{\cal R}_{ii} =  \frac{4 \pi}{3} \rho \, N \frac{\Omega_t}{4\pi} \ \left( \frac{{\cal L}_\nu}{4 \pi  {\cal S}} \right)^{3/2}.
\end{eqnarray}
The expected detection rate using a fly's eye configuration is a factor $N^{1/4}$ higher than that of the collimated configuration.  

This demonstrates the shortcomings of this particular FoM as the overall metric of a fast transients survey, particularly for large-$N$ arrays such as the SKA.  The discrepancy between the two survey metrics stems from the fact that the survey speed is proportional to $S_{0}^{-2}$, while the detection rate has a softer dependence of $S_{0}^{-3/2}$; in the former case there is no advantage in survey speed to be gained by trading collecting area per FoV for a commensurate sensitivity decrease.  In the latter case,
an incoherent survey whose sensitivity reaches a factor of $N^{1/2}$ deeper in sensitivity yields $N^{3/4}$ more objects for each FoV, but its FoV is $N$ times less than that of a fly's eye experiment, yielding a nett penalty of $N^{1/4}$ in detection rate over the fly's eye mode.  Fundamentally, the difference between these two metrics stems from the fact that instantaneous sensitivity can be traded against integration time for a survey of steady sources, but not for a transients survey.  This highlights the point, relevant to surveys for fast transients, that $V_{\max}$ is independent of dwell time per telescope pointing, as long as this time considerably exceeds the short duration of the events themselves.


An additional limitation of the FoM when investigating survey strategy is that it does not permit further exploration of its dependence on the detailed properties of the population, such as its luminosity distribution and its apparent angular distribution on the sky.  The latter is an important consideration for Galactic transients, as discussed in \S\ref{sec:galactic} below.

\section{Extragalactic Transients} \label{sec:extragalactic}

We estimate the event rate of a putative population of transient objects as a function of limiting flux density.  We consider a population of transients distributed homogeneously through space with a power-law luminosity distribution of the type described in eq.\,(\ref{Ldistn}).    There are $\rho_{\cal L} d{\cal L}$ objects in the luminosity range ${\cal L}$ to ${\cal L}+d{\cal L}$ per unit volume.  These objects can be detected out to a limiting volume $V_{\rm max} = (4 \pi/3) D_{\rm max}^3 = (4 \pi/3) ({\cal L}/4 \pi S_0)^{3/2}$, where $S_0$ is the minimum detectable flux density of a transient in the survey\footnote{For the purposes of simplicity we ignore effects associated with the curvature of spacetime, and thus approximate spacetime as Euclidean out to the edge of the survey volume.  This is sufficient for surveys for objects at $z \ll 1$.  The generalization to larger distances complicates the algebra but is straightforward, and involves replacing the distance with the luminosity distance and the volume of the shell with a comoving volume (see, e.g., Nemiroff 2003).}.  

The number of objects detectable in the luminosity range ${\cal L}$ to ${\cal L}+d{\cal L}$ is, 
\begin{eqnarray}
\rho_{\cal L} d{\cal L} \frac{4\pi}{3} \left( \frac{{\cal L}}{4 \pi S_0}\right)^{3/2} = \frac{\rho_0}{K} {\cal L}^{-\alpha} \frac{4\pi}{3} 
\left( \frac{{\cal L}}{4 \pi S_0} \right)^{3/2} d{\cal L} .
\end{eqnarray}
The total event rate, for a finite FoV and integrated over all luminosities is,
\begin{eqnarray}
{\cal R}_{\rm sens-bound} &=& \frac{4 \pi \rho_0}{3 K} \frac{\Omega}{4 \pi} \int_{{\cal L}_{\rm min}}^{{\cal L}_{\rm max}} {\cal L}^{-\alpha} \left( \frac{L}{4 \pi S_0} \right)^{3/2} d{\cal L} \nonumber \\
&=& \frac{\rho_0 S_0^{-3/2}}{3 \sqrt{4 \pi}} \frac{\Omega}{4\pi} \left( \frac{1-\alpha}{5/2 -\alpha} \right) \frac{{\cal L}_{\rm max}^{5/2-\alpha} - {\cal L}_{\rm min}^{5/2-\alpha}}{{\cal L}_{\rm max}^{1-\alpha} - {\cal L}_{\rm min}^{1-\alpha}}, \quad \alpha \neq 5/2,
\label{SensBounded}
\end{eqnarray}
where the subscript on ${\cal R}$ refers to the fact that the survey is bounded purely by the sensitivity of the observations.  For the purposes of the following section, in which the detection rate varies with sky position, it is convenient to define the detection rate in the region of sky per small solid angle $d \Omega$ as $R_{\rm sens-bound}$, such that 
\begin{eqnarray}
{\cal R}_{\rm sens-bound} = \int_{\rm FoV} R_{\rm sens-bound} d\Omega.
\end{eqnarray}

Writing ${\cal L}_{\rm max}/{\cal L}_{\rm min} = \beta \gg 1$, the expression for the detection rate divides into four cases: (i) $\alpha < 1$, (ii) $1 < \alpha < 5/2$, and (iii) $\alpha=5/2$ and (iv) $\alpha > 5/2$.   In the limit $\beta \gg 1$ the detection rate reduces to the following form:
\begin{eqnarray}
{\cal R} &=&  \frac{\rho_0 \Omega}{24 \pi^{3/2}}   
\left( \frac{{\cal L}_{\max}}{S_0} \right)^{3/2} g(\alpha, \beta), \label{RformEq}
\end{eqnarray}
where we define,
\begin{eqnarray}
g(\alpha, \beta) = \left\vert \frac{1-\alpha}{5/2-\alpha} \right \vert \left\{ \begin{array}{ll}
1, & \alpha < 1, \\
\beta^{1-\alpha}, & 1< \alpha < 5/2, \\
(5/2-\alpha) \beta^{-3/2} \ln \beta & \alpha = 5/2, \\
\beta^{-3/2}, & \alpha > 5/2. \\
\end{array} \right.
\end{eqnarray}
It is convenient to express the rate in terms of some fiducial numbers for the luminosity and event rate,
\begin{eqnarray}
{\cal R} &=& 2.3 \times 10^{-6} \, g(\alpha,\beta) \,  
\left( \frac{\rho_0}{1 \,{\rm event\,\,s}^{-1}{\rm Gpc}^{-3}} \right) 
\left( \frac{\Omega}{1\,{\rm deg}^2} \right) 
\left( \frac{S_0}{1\,{\rm mJy}} \right)^{-3/2} 
\left( \frac{{\cal L}_{\max} }{1\,{\rm mJy\,Gpc}^{2}} \right)^{3/2} \,\,{\rm events\,s}^{-1}.   \nonumber \\ 
\end{eqnarray}

\subsection{An example: ASKAP}
To put this in perspective, we consider some numbers relevant to the 36-element ASKAP interferometer.  The FoV for a single-pixel fully coherent detection mode (in which all the visibilities are combined and searched) is $\sim \pi(\lambda/d)^2$, equal to $1.3 \times 10^{-5}$\,deg$^2$~at $\lambda 21\,$cm with a baseline of $d=6$\,km.  The FoV for an incoherent collimated survey mode is 30\,deg$^2$, and is $36\times 30\,$deg$^2$~for the fly's-eye mode of operation.  If the instantaneous minimum detectable sensitivity is ${\cal S}$ in the coherent mode for a given integration time, it is $\sqrt{36}\, {\cal S}$ in the incoherent mode and $36\,{\cal S}$ in the fly's-eye mode.  Thus the detection rates for the various modes are,
\begin{eqnarray}
{\cal R} &=& g(\alpha,\beta) \left\{ \begin{array}{ll} 
9.1 \times 10^{-16}, & \hbox{coherent}\\
1.5 \times 10^{-10}, & \hbox{incoherent}\\
3.6 \times 10^{-10}, & \hbox {fly's-eye} \\
\end{array} \right\} 
  \left( \frac{\rho_0}{1 \,{\rm event\,\,s}^{-1}{\rm Gpc}^{-3}} \right) 
\left( \frac{L_{\max} }{1\,{\rm mJy\,Gpc}^{2}} \right)^{3/2}  \left( \frac{\cal S}{1\,{\rm Jy}} \right)^{-3/2} 
\nonumber \\ &\null& 
\qquad \qquad \qquad \qquad \qquad \qquad \qquad \qquad \qquad \qquad 
\qquad \qquad \qquad \qquad \qquad {\rm events\,s}^{-1}.
\end{eqnarray}
An important point is that the detection rate of the fly's-eye mode is superior to that of the incoherent detection mode, irrespective of the actual cutoff in the luminosity distribution.  Which is to say that, even if the transients are dominated by the low-luminosity end,  the incoherent mode, even though it probes weaker events relative to the fly's-eye mode, offers no advantage in detection rate.

\subsection{Distance distribution of detected transients}
We employ the foregoing formalism to calculate the distance distribution of transients in the survey.  The distribution of distances to these transients is an important consideration because it dictates the amount of detection and post-processing power required.  We would like to know, for example, whether the detection rate is dominated by near or far objects, and thus the typical range of dispersion measures that our detection hardware should encompass.  The distance also plays an important role in determining how much scattering the radiation is subject to, and this can in principle pose serious limitations to the detectability of extremely short-duration transients.  For the moment we shall assume that the objects we seek to detect are of sufficient duration that temporal smearing caused by intergalactic scattering is negligible.  This is a good assumption for most lines of sight, since the intergalactic medium likely only makes a modest (though perhaps not unmeasurable) contribution to pulse smearing, and the``lever-arm'' effect associated with the optics of temporal smearing renders the contribution of scattering material in our Galaxy's interstellar medium small.

We compute the fraction of events, ${\cal F}(D)$, above a flux density limit $S_{\rm min}$ at a distance $D$.   The flux density of an object of luminosity $L_0$ is $L_0/4 \pi D^2$, and this will be detectable if $L_0 > 4 \pi D^2 S_0$.  Thus the number rate of objects detectable on a shell between distances $D$ and $D+dD$, given by the product of the volume encompassed by the volume of the shell that is visible within the FoV $\Omega$ and the local space density rate of events whose luminosities fall in the range of detectability, is,
\begin{eqnarray}
{\cal N}(D) dD = 4 \pi D^2 dD \frac{\Omega}{4\pi} \int_{\max[4\pi D^2 S_0,{\cal L}_{\rm min}]}^{\max[{\cal L}_{\rm max},4 \pi D^2 S_0]} \rho_{\cal L}({\cal L}) d{\cal L}. \label{Nintegral}
\end{eqnarray}
The corresponding fraction of objects detectable is ${\cal F}(D) = {\cal N}(D)/{\cal R}$.  

The integral splits into three cases, depending on whether (i) our sensitivity is so good that we detect all objects at a given distance ($4 \pi D^2 S_0 < {\cal L}_{\rm min}$), (ii) our sensitivity is intermediate, so we only detect some of the objects (${\cal L}_{\rm min} < 4 \pi D^2 S_0 < {\cal L}_{\rm max}$), or (iii) it is so poor that we do not even detect the brightest objects (${\cal L}_{\rm max} < 4 \pi D^2 S_0$).  One therefore has, for $\alpha \neq 1$,
\begin{eqnarray}
{\cal N}_{\rm sens-bound}(D) dD =  \frac{\Omega \rho_0 D^2 dD }{(1-\alpha) K} {\cal L}_{\max}^{1-\alpha} \left\{ \begin{array}{ll}
1 - \beta^{\alpha-1}, & \beta^{-1} > 4 \pi D^2 S_0/{\cal L}_{\max}, \\
1 - (4 \pi D^2 S_0/{\cal L}_{\max})^{1-\alpha}, & \beta^{-1} < 4 \pi D^2 S_0/{\cal L}_{\max} < 1, \\
0, & 1 < 4 \pi D^2 S_0/{\cal L}_{\max} , \\ 
\end{array} \right. \label{ExtragalacticDistanceDistn} 
\end{eqnarray}
The shape of the distribution depends only on the ratio $S_0/{\cal L}_{\max}$, $\beta$ and $\alpha$.  Again, for the purposes of the following section, it is convenient to make the auxiliary definition $N_{\rm sens-bound}$ as the event rate per unit distance per unit solid angle,
\begin{eqnarray}
{\cal N}_{\rm sens-bound}(D) = \int_{\rm FoV} N_{\rm sens-bound} d\Omega .
\end{eqnarray}

\begin{figure}[htb!]
\begin{center}
\epsfig{file=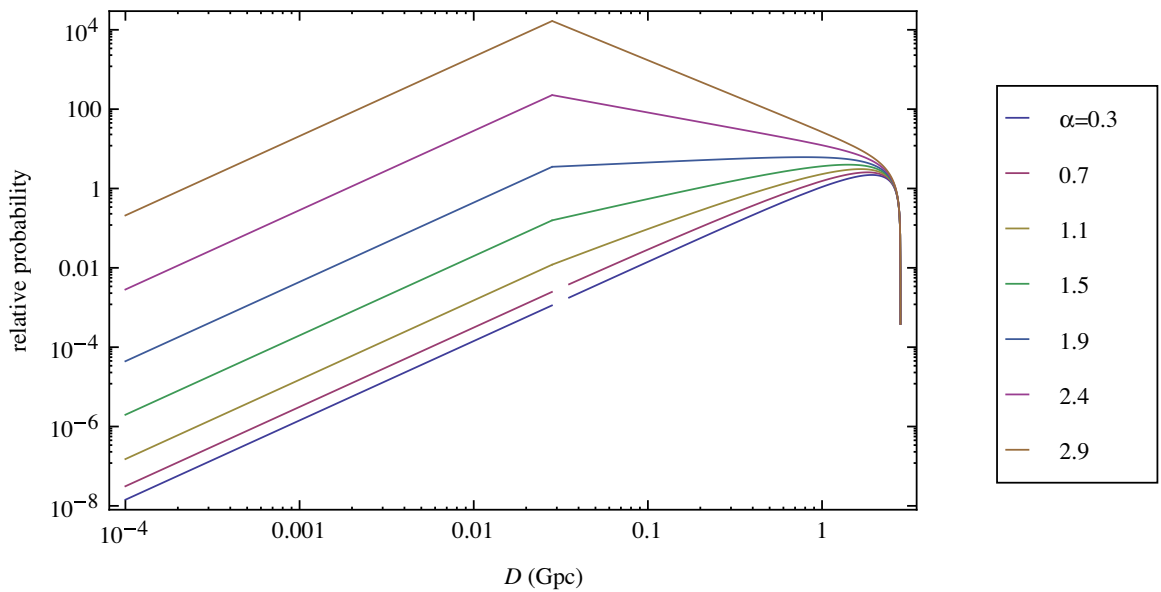,scale=0.9}
\caption{A plot of the relative probability of detection, ${\cal N}(D) [\Omega {\cal L}_{\max}^{1-\alpha}/(1-\alpha) K]^{-1}$, for various values of $\alpha$, with $\beta=10^4$ and $S_0/{\cal L}_{\max} = 0.01\,$Gpc$^{-2}$.  The distribution rises as $D^2$ until the first break at $({\cal L}_{\max} /4 \pi \beta S_0)^{1/2}$, and then continues proportional to $D^{4-2\alpha}$ until the second turnover occurs at $({\cal L}_{\max}/4 \pi S_0)^{1/2}$.
} \label{DistnFig}
\end{center}
\end{figure}

A plot of ${\cal N}(D)$ is shown in Fig.\,\ref{DistnFig} for luminosity distributions of various indices.  For flat luminosity distributions the event rate is dominated by objects at the extreme range of the survey, but as the luminosity distribution steepens the event rate is increasingly dominated by events closer in.  For luminosity distributions shallower than $\alpha=2$ the event rate distribution continues rising until $D=({\cal L}_{\max}/4 \pi S_0)^{1/2}$, whereas for steeper luminosity distributions, $\alpha>2$, the event rate distribution peaks at $D = ({\cal L}_{\min} /4 \pi S_0)^{1/2}$.

For completeness, we also give the result for the special case $\alpha =1$,
\begin{eqnarray}
{\cal N}(D) dD =  \Omega \rho_0 D^2 dD \left\{ \begin{array}{ll}
1, & {\cal L}_{\min} > 4 \pi D^2 S_0, \\
(\ln \beta )^{-1} \ln \left( \frac{{\cal L}_{\max}}{4 \pi D^2 S_0} \right), & {\cal L}_{\min} < 4 \pi D^2 S_0 < {\cal L}_{\max}, \\
0, & {\cal L}_{\max} < 4 \pi D^2 S_0. \\ 
\end{array} \right. 
\end{eqnarray}

\subsection{Lognormal distribution}
One may wonder to what extent the foregoing results depend on the specific assumption that the luminosity distribution follows a power law.  To this end, we present the results for a luminosity function following a lognormal distribution,
\begin{eqnarray}
{\rho_{\cal L}}_{\rm lognorm} d{\cal L} = \frac{\rho_0}{\sqrt{2 \pi \sigma^2 {\cal L}^2}}  \exp \left[ - \frac{(\ln {\cal L} - \mu)^2}{2 \sigma^2} \right] .
\end{eqnarray} 
The only free parameters of the model are the location of the peak of the distribution, parameterized by $\mu$, and its width, parameterized by $\sigma$.  The mean luminosity is $\langle {\cal L} \rangle = \exp( \mu + \sigma^2/2)$ and its variance is given by ${\rm var}({\cal L}) \equiv \langle ({\cal L}-\langle {\cal L} \rangle)^2 \rangle = (e^{\sigma^2/2} -1) \exp (2 \mu + \sigma^2)$. The parameter $\sigma^2$ can be interpreted using the relation 
\begin{eqnarray}
1+\frac{{\rm var}({\cal L})}{\langle {\cal L} \rangle^2} = \exp \sigma^2. \label{sigmaDefn}
\end{eqnarray}  This distribution possesses the virtue that no lower or upper luminosity cutoffs need by imposed.
The lognormal distribution is a common alternative to the power law luminosity distribution, particularly for the case of pulsars.  For instance, Faucher-Gigu\`ere \& Kaspi (2006) show that the pulsar luminosity distribution can be modelled well by such a distribution.

The expected event rate is 
\begin{eqnarray}
{\cal R} &=& \frac{4 \pi }{3} \frac{ \Omega}{4 \pi} \frac{\rho_0}{\sqrt{2 \pi \sigma^2 L^2}} \int_0^\infty d{\cal L} \left( \frac{\cal L}{4 \pi S_0}\right)^{3/2}  \exp \left[ - \frac{(\ln {\cal L} - \mu)^2}{2 \sigma^2} \right] \nonumber \\
&=&  \frac{ \Omega \rho_0}{24 \,\pi^{3/2}} S_0^{-3/2} \exp \left[ \frac{3 \mu}{2} + \frac{9 \sigma^2}{8} \right] .
\end{eqnarray}
It is instructive to represent this result in terms of the mean and variance of the luminosity, so that it may be directly compared with the corresponding result for a power law distribution (viz. eq.(\ref{RformEq})),
\begin{eqnarray}
{\cal R} = 
\frac{ \Omega \rho_0}{24 \,\pi^{3/2}} \left(\frac{\langle {\cal L} \rangle}{S_0} \right)^{3/2} \left(  1+ \frac{{\rm var}({\cal L})}{\langle {\cal L} \rangle^2} \right)^{3/8}.
\end{eqnarray}
In particular, we see that this event rate expression is similar to eq.(\ref{RformEq}), and can be cast in exactly the same form with the replacements ${\cal L}_{\max} \rightarrow \langle {\cal L} \rangle$ and $g(\alpha,\beta) \rightarrow \left[  1+ {\rm var}({\cal L})/\langle {\cal L} \rangle^2 \right]^{3/8}$.

\section{Galactic Transients} \label{sec:galactic}
\begin{figure}[htb!]
\begin{center}
\epsfig{file=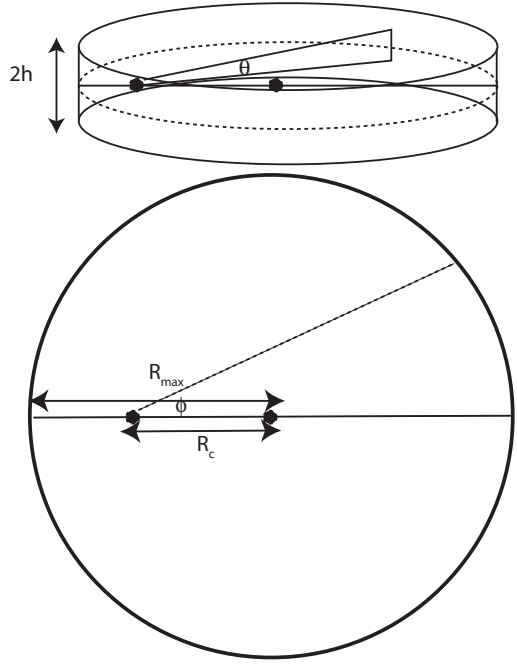,scale=0.7}
\caption{We consider a model of the Galaxy in which transients are confined within the volume of a flattened cylinder.} \label{figGalaxyGeometry}
\end{center}
\end{figure}

In this section we repeat the above analysis for a population of transients confined to our Galaxy.
The calculation is less straightforward because the volume $V_{\max}$ is a complicated function that depends on the geometry of the Galaxy.  Moreover, since the observer is not situated at the Galactic Centre, the survey volume and thus the rate is also direction-dependent.  For the purposes of simplicity we model the Galaxy as a flattened cylinder of radius $R_{\max}$ and height $2h$, with the observer (i.e. the Earth) located on the midplane of the Galaxy a distance $R_c$ from the centre of the Galaxy (see Fig.\,\ref{figGalaxyGeometry}).  We assume the population of transients to be situated homogeneously within this volume.  Although this assumption may appear crude, it is sufficient to explore the problem to the level of approximation we are interested in, and it is likely more than sufficient given the uncertainties in the demographics of any putative transients populations. 

\begin{figure}[htb!]
\begin{center}
\epsfig{file=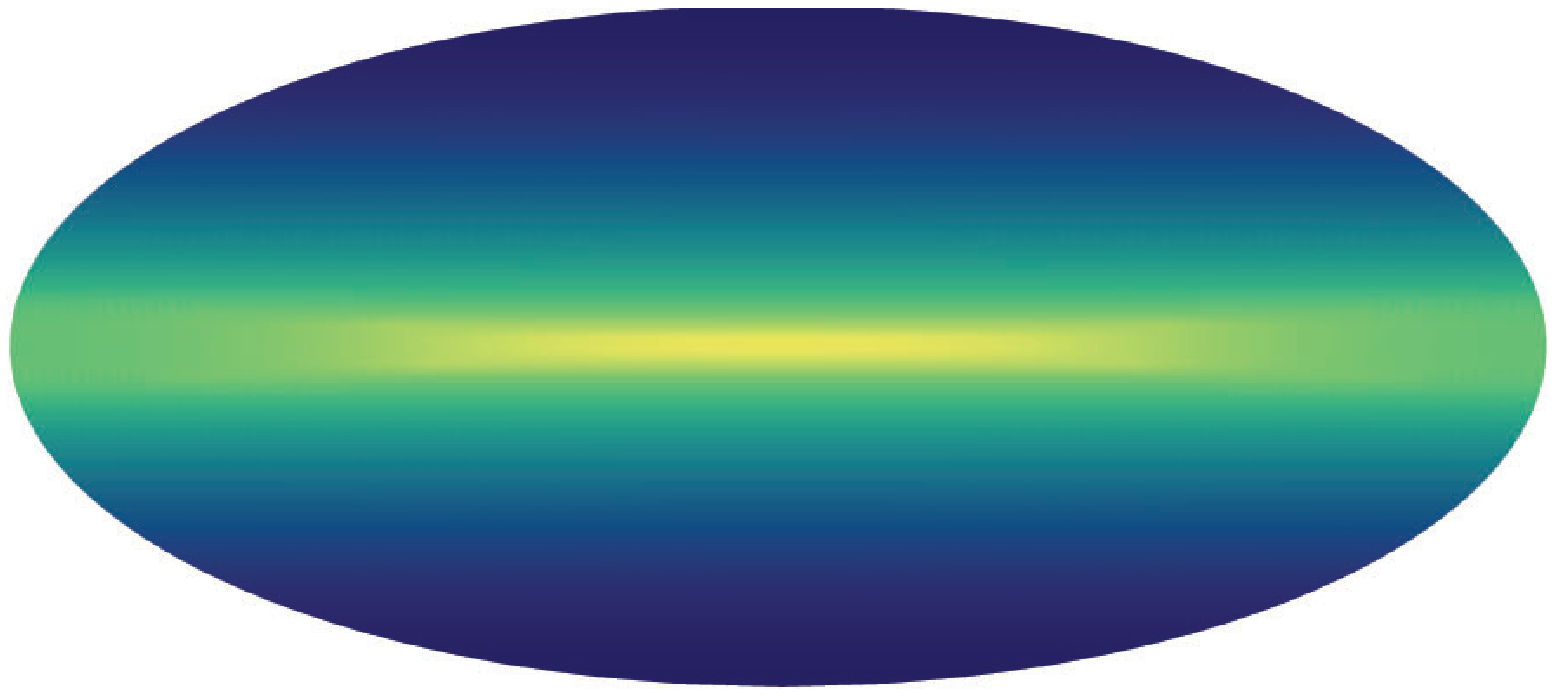,scale=1.0} \\
\epsfig{file=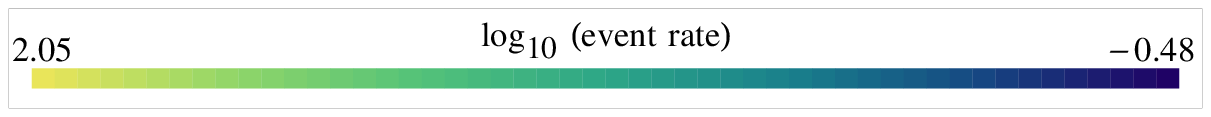,scale=0.7} \\
\caption{The normalized event rate in the Galaxy ${\cal R}/\rho_0 d\Omega$ as a function of Galactic position for ${\cal L}_{\min} = 0.1$Jy\,kpc$^2$, ${\cal L}_{\max} = 100$Jy\,kpc$^{2}$, $\alpha = 1.5$, $R_{\max}=15\,$kpc, $R_c=8\,$kpc, $h=1.0\,$kpc.  The event rate is colour coded according to the log$_{10}$ of the event rate.  Here $S_0 = 10\,$mJy.     
} \label{figGalacticEvents}
\end{center}
\end{figure}

Now, if the line of sight is at low Galactic latitude, the line of sight will intersect the thin edge of the cylinder.  The locus of the edge of the Galaxy is $x^2 + y^2 = R_{\max}^2$, and the locus of the line of sight in the $x-y$ plane is $(r \cos \phi - R_c, r \sin \phi)$.  These two lines intersect at,
\begin{eqnarray}
r_{\pm} = R_c \cos \phi \pm \sqrt{R_{\max}^2 - R_c^2 \sin^2 \phi}.
\end{eqnarray}
For the geometry given in Fig.\,\ref{figGalaxyGeometry} only the positive root, $r_+$, is of interest.
This solution is only relevant if the elevation is sufficiently low that $|r_+ \tan \theta | < h$.  The total distance is then
\begin{eqnarray}
D_{\max} = \sqrt{r_+^2 + r_+^2 \tan \theta^2 } = r_+ \sec \theta, \qquad |r_+ \tan \theta | < h.
\end{eqnarray}
Otherwise, the line of sight intersects the edge of the Galaxy on the face of the cylinder and we have
\begin{eqnarray}
D_{\max} = \frac{h}{|\sin \theta|}, \qquad  |r_+ \tan \theta | \geq h.
\end{eqnarray}
In summary, we have $V_{\max} = (\Omega /3) D_{\max}^3$ with, 
\begin{eqnarray}
D_{\max}(\theta,\phi) = \left\{ \begin{array}{ll}
\left( \frac{{\cal L}}{4 \pi S_0} \right)^{1/2}, &  
	\left( \frac{{\cal L}}{4 \pi S_0} \right)^{1/2} < d(\theta,\phi) \\
d(\theta,\phi), & \left( \frac{{\cal L}}{4 \pi S_0} \right)^{1/2} > d (\theta,\phi) \\ 
\end{array} \right.
\end{eqnarray}
where the distance out to the edge of the Galaxy along the line of sight is,
\begin{eqnarray}
d(\theta,\phi) = \left\{ \begin{array}{ll} 
		r_+ \sec \theta, & h > | r_+ \tan \theta |, \\
		h/|\sin \theta|, &  h < | r_+ \tan \theta |. \\
		\end{array} \right.
\end{eqnarray}
We make the approximation that the FoV, $\Omega$, is sufficiently narrow that the survey volume is a narrow cone of volume $V_{\max} = (4\pi/3)  D_{\max}^3 (\Omega/4 \pi)$ (i.e. that there is no appreciable variation in $D_{\max}$ across the FoV).  We define ${\cal R}_\Omega$ as the differential event rate per solid angle such that this quantity, when integrated across the field of view, yields the total event rate observed over the FoV:
\begin{eqnarray}
{\cal R} = \int_{\rm FoV} R_\Omega \, d\Omega.
\end{eqnarray}
The distance $D_{\max}$ is a function of Galactic latitude and longitude, $\theta$ and $\phi$ respectively.  There are two cases to consider: (i) low sensitivity in which $D_{\max} = ({\cal L}/4 \pi S_0)^{1/2}$ is less than the distance from the observer to the edge of the Galaxy along the direction of interest and (ii) high sensitivity in which the observations probe out to the edge of the Galaxy along the direction of interest.  We integrate this over the luminosity function to derive the total rate along the line of sight,
\begin{eqnarray}
{\cal R}_\Omega (\theta,\phi) d\Omega = \frac{\rho_0 d\Omega}{3 K} \int_{{\cal L}_{\min}}^{{\cal L}_{\max}} d{\cal L} \, {\cal L}^{-\alpha} 
 \left\{ \begin{array}{ll}
\left( \frac{{\cal L}}{4 \pi S_0} \right)^{3/2}, &  
	{\cal L}_0 > {\cal L}, \\
d^3(\theta,\phi), & {\cal L}_0 < {\cal L} . \\ 
\end{array} \right.
\end{eqnarray}
The quantity ${\cal L}_0 \equiv 4 \pi S_0 d^2(\theta,\phi)$ is the luminosity needed to detect the object out to the edge of the Galaxy along that line of sight.  This integral splits into three cases (i) low sensitivity observations, where all the objects are below the luminosity required for the observations to probe all the objects to the edge of the Galaxy: $ {\cal L}_{\min} < {\cal L} < {\cal L}_{\max} < {\cal L}_0$, 
(ii) intermediate sensitivity observations, where some high-luminosity objects are observable to the edge of the Galaxy (${\cal L}_{\min} < {\cal L}_0 < {\cal L} < {\cal L}_{\max}$), while the low-luminosity objects are only detectable along a portion of the line of sight (${\cal L}_{\min} < {\cal L} < {\cal L}_0< {\cal L}_{\max}$) and (iii) high sensitivity observations, where all the objects exceed the luminosity required for them to be observable to the edge of the Galaxy ${\cal L}_0 < {\cal L}_{\min} < {\cal L} < {\cal L}_{\max}$.  The expression for the detection rate therefore divides into three categories: 
\begin{eqnarray}
{\cal R}_\Omega d \Omega &=& \frac{\rho_0 d \Omega}{3 K} \left\{ \begin{array}{ll}
\int_{{\cal L}_{\min}}^{{\cal L}_{\max}} d{\cal L} \, {\cal L}^{-\alpha} \left( \frac{{\cal L}}{4 \pi S_0} \right)^{3/2}, & {\cal L}_0 > {\cal L}_{\max}, \\
\int_{{\cal L}_{\min}}^{L_0} d{\cal L} \, {\cal L}^{-\alpha} \left( \frac{{\cal L}}{4 \pi S_0} \right)^{3/2} +  \int_{{\cal L}_0}^{{\cal L}_{\max}} d{\cal L} \, {\cal L}^{-\alpha} d^3(\theta,\phi),  & {\cal L}_{\min} < {\cal L}_0 < {\cal L}_{\max}, \\
\int_{{\cal L}_{\min}}^{{\cal L}_{\max}} dL \, {\cal L}^{-\alpha} d^3(\theta,\phi), & {\cal L}_0 < {\cal L}_{\min}. \\ 
\end{array} \right. \nonumber \\
&=& \frac{\rho_0 d\Omega}{3 K} \left\{ \begin{array}{ll}
\left( 4 \pi S_0 \right)^{-3/2} \frac{1}{5/2-\alpha} \left( {\cal L}_{\max}^{5/2-\alpha} - {\cal L}_{\min} ^{5/2-\alpha} \right) , & {\cal L}_0 > {\cal L}_{\max} \\
\left( 4 \pi S_0 \right)^{-3/2} \frac{1}{5/2-\alpha} \left( {\cal L}_{0}^{5/2-\alpha} - {\cal L}_{\min} ^{5/2-\alpha} \right) + \frac{d^3(\theta,\phi)}{1-\alpha} \left( {\cal L}_{\max}^{1-\alpha} - {\cal L}_{0}^{1-\alpha} \right), &   {\cal L}_{\min} < {\cal L}_0 < {\cal L}_{\max} \\
\frac{d^3(\theta,\phi)}{1-\alpha} \left( {\cal L}_{\max}^{1-\alpha} - {\cal L}_{\min}^{1-\alpha} \right), & {\cal L}_0 < {\cal L}_{\min}. \\
\end{array} \right.  \nonumber \\
\end{eqnarray}
The expression for the event rate in the case ${\cal L}_0 > {\cal L}_{\max}$ is, of course, identical to the rate, $R_{\rm sens-bound}$ derived in eq.(\ref{SensBounded}) for a homogeneous sensitivity-bounded (as opposed to a volume-bounded) survey.  In the high sensitivity case the rate is completely volume-bounded, and the detection rate depends only on the number of objects visible within the field of view: $R=\rho_0 \Omega d^3(\theta,\phi)/3$.  In the intermediate case, ${\cal L}_{\min} < {\cal L}_0 < {\cal L}_{\max}$, the survey is sensitivity-limited at low luminosities but volume-limited at high luminosities.
\begin{eqnarray}
{\cal R}_\Omega d \Omega &=& \left\{ \begin{array}{ll}
R_{\rm sens-bound} d\Omega, & {\cal L}_0 > {\cal L}_{\max}, \\
R_{\rm sens-bound} \left( \frac{ {\cal L}_{0}^{5/2-\alpha} - {\cal L}_{\min}^{5/2-\alpha}}{ {\cal L}_{\max}^{5/2-\alpha} - {\cal L}_{\min}^{5/2-\alpha}} \right) d\Omega  + [\rho_0 \, d \Omega \, d^3(\theta,\phi)/3] \left( \frac{ {\cal L}_{\max}^{1-\alpha} - {\cal L}_{0}^{1-\alpha}}{ {\cal L}_{\max}^{1-\alpha} - {\cal L}_{\min}^{1-\alpha}} \right), &   {\cal L}_{\min} < {\cal L}_0 < {\cal L}_{\max}, \\
\rho_0 \, d\Omega \, d^3(\theta,\phi)/3, & {\cal L}_0 < {\cal L}_{\min}. \\
\end{array} \right. \nonumber \\ \label{GalaxyRate} 
\end{eqnarray}
 A plot of the event rate distribution across the Galaxy is shown in Fig.\,\ref{figGalacticEvents} for some fiducial survey and population parameters.  The ``star-shaped'' pattern associated with a high event rate emanating from the Galactic Centre and extending outwards a few degrees either side of the Galactic plane reflects the fact that, in the model of the Galaxy geometry considered here, the longest sight lines through the Galaxy are those that intersect the vertices of the cylinder wall with its two faces (see Fig.\,\ref{figGalaxyGeometry}).  However, the simple generic form of eq.\,(\ref{GalaxyRate}) shows that it is straightforward to substitute an alternate model for the geometry of the Galaxy or distribution of the transient population by replacing the functional form of $d(\theta,\phi)$ if desired.  However, this is a minor concern given the large uncertainties in the quantifying the true spatial distribution of any putative population of transients within the Galaxy.

\subsubsection{Results for a lognormal distribution}
For the sake of completeness, we quote the corresponding Galactic event rate when the luminosity follows a lognormal distribution.  The absence of cutoffs, ${\cal L}_{\rm min}$ and ${\cal L}_{\max}$, in the distribution simplifies the algebra.  The event rate is comprised of objects from luminosity $0$ to ${\cal L}_0$ which are too faint to detect up to the boundary of the Galaxy, and objects that are sufficiently luminous that they are all detected out to Galactic edge,
\begin{eqnarray}
{\cal R}_\Omega d \Omega &=& \frac{\rho_0 d\Omega}{3} \left[ \int_0^{{\cal L}_0} \left( \frac{{\cal L}}{4 \pi S_0} \right)^{3/2} {\rho_{\cal L}}_{\rm lognorm} d{\cal L} + \int_{{\cal L}_0}^\infty d^3(\theta,\phi) {\rho_{\cal L}}_{\rm lognorm} d{\cal L} \right] \nonumber \\
&=& \frac{\rho_0 d\Omega }{48 \pi^{3/2}} \left(\frac{\langle {\cal L} \rangle}{S_0} \right)^{3/2} \left(  1+ \frac{{\rm var}({\cal L})}{\langle {\cal L} \rangle^2} \right)^{3/8} \left[  1 + {\rm erf} \left( \frac{-2\mu - 3 \sigma^2 +  2\ln {\cal L}_0}{2 \sqrt{2}\sigma} \right) \right]  \nonumber \\
&\null&  \qquad + \frac{\rho_0 d\Omega}{6} d^3(\theta,\phi) \left[ 1 + {\rm erf} \left(\frac{\mu - \ln {\cal L}_0}{\sqrt{2} \sigma} \right) \right].  \nonumber \\
\end{eqnarray}
This result is not employed in any subsequent calculations, but it does demonstrate the similarity between the expected event rate for power law and lognormal luminosity distributions.  It resembles the expression for the event rate in the case ${\cal L}_{\min} < {\cal L}_0 < {\cal L}_{\max}$ for the power law distribution, with ${\cal L}_{\min}$ set to zero and with no bound on the upper luminosity.

\subsection{Number of transients as a function of distance}
The distribution of transient events as a function of distance is readily derived based on the simple form of the transients rate, as expressed in eq.\,(\ref{GalaxyRate}).  In the sensitivity-bound regime, ${\cal L}_0 > {\cal L}_{\max}$, the distribution of event distances is identical to that found in the unbounded case, in eq.\,(\ref{ExtragalacticDistanceDistn}).   In the opposite extreme, in the volume-bound regime in which ${\cal L}_0 < {\cal L}_{\min}$, the survey detects all transients that are visible inside the cone of opening solid angle $\Omega$.  The number of objects per differential solid angle $d \Omega$ detected between a distance $D$ and $D+dD$ from the observer is simply the number of objects on a thin slice of cone with opening angle $d\Omega$, 
\begin{eqnarray}
{\cal N}_{\rm \Omega\, vol-bound}(D) dD d\Omega = \rho_0  D^2 \, d\Omega\, dD, \qquad D \leq d(\theta,\phi).
\end{eqnarray}
The intermediate regime, ${\cal L}_{\min} < {\cal L}_0 < {\cal L}_{\max}$ contains a mixture of these two solutions.  At low luminosities the distance distribution is governed by the solution for sources that are sensitivity bound, whereas at high luminosities the distance distribution is governed by the solution in the volume-bound regime.  

In summary, the distance distribution of transient rates is given by,
\begin{eqnarray}
{\cal N}_{\Omega} (D) dD d \Omega = dD  d\Omega \left\{ \begin{array}{ll}
N_{\rm sens-bound}(D) , & {\cal L}_0 > {\cal L}_{\max}, \\
\left( \frac{ {\cal L}_{0}^{5/2-\alpha} - {\cal L}_{\min}^{5/2-\alpha}}{ {\cal L}_{\max}^{5/2-\alpha} - {\cal L}_{\min}^{5/2-\alpha}} \right)  N_{\rm sens-bound}(D)  & \null \\
\qquad \qquad \quad + \rho_0 \, D^2  \left( \frac{ {\cal L}_{\max}^{1-\alpha} - {\cal L}_{0}^{1-\alpha}}{ {\cal L}_{\max}^{1-\alpha} - {\cal L}_{\min}^{1-\alpha}} \right), & {\cal L}_{\min} < {\cal L}_0 < {\cal L}_{\max},\,\, D < d(\theta,\phi), \\
\rho_0 \, D^2, & {\cal L}_0 < {\cal L}_{\min},\,\, D < d(\theta,\phi). \\
\end{array} 
\right. \label{NdistanceDistn}
\end{eqnarray}
The ability to express the event rate as a function of distance and Galactic co-ordinate permits a number of obvious generalizations.  The first is that one can relax the assumption of a uniform density of transient progenitors in the Galaxy.  A non-uniform intrinsic event rate is treated by replacing the constant $\rho_0$ with a function $\rho_0(\theta, \phi, D)$ which describes the event rate density along each line of sight as a function of distance $D$ from the observer.  Such a generalization is useful when examining known classes of transients, such as RRATs, for which the density of the progenitor population, such as neutron stars, is known (see, e.g., Ofek et al. 2010).   We retain the assumption of fixed event rate density throughout the remainder of the present work rather than adopt a specific assumption for some hypothetical population of transients.  Strategies for surveys of short-timescale emission from one specific class of objects, neutron stars, have been studied extensively elsewhere (Smits et al.\,2009).  The second obvious generalization is to include the distance and direction-dependent effects of scattering by inhomogeneities in the ionized Interstellar Medium.  These are considered below.

\subsection{The effect of temporal smearing caused by interstellar scattering} \label{sec:scat}
The foregoing calculations omit the effect of interstellar scattering.  This is relevant to any signal whose duration is smaller than or comparable to the pulse broadening time set by multi-path propagation through the turbulent interstellar medium.  For a pulse of intrinsic duration $\Delta t$ that is broadened to a duration $T \approx \sqrt{\tau^2 + \Delta t^2}$, where $\tau$ is the scattering broadening timescale, the pulse flux density decreases by a factor of $\approx T/\Delta t$, while the sensitivity of the integration is increased by a factor $\approx \sqrt{T/\Delta t}$ over an integration over a time $\Delta t$.  Thus the overall sensitivity loss is a factor $\sqrt{T/\Delta t}$.  

We incorporate the sensitivity loss inherent to interstellar temporal smearing by including a distance-dependent modification to the minimum detectable luminosity in our survey $L_0 = 4 \pi D^2 S_0 f(D)$, where the function $f(D)$ embodies the loss of sensitivity caused by temporal smearing.  One has,
\begin{eqnarray}
f(D) = \sqrt{T/\Delta t} = \left( \frac{\sqrt{\tau^2(D,\theta,\phi) + \Delta t^2}}{\Delta t} \right)^{1/2},
\end{eqnarray}
where we have written $\tau(D,\theta,\phi)$ as an explicit function of the distance and the direction of the line of sight through the Galaxy.  Since $\tau$ depends on the detail of structures in our Galaxy, it must be computed numerically for each combination of $D$, $\theta$ and $\phi$.  We use the NE2001 scattering model (Cordes \& Lazio 2002) to compute $\tau$ using the code available at {\tt http://rsd-www.nrl.navy.mil/7213/lazio/ne\_model/}.  This code yields the temporal smearing time at a frequency of 1\,GHz, but readily scales to other frequencies under the assumption that the turbulence follows a Kolmogorov spectrum, so that $\tau \propto \nu^{-4.4}$.  

Although the assumption of Kolmogorov turbulence is broadly valid (Amstrong, Rickett \& Spangler 1995) and underpins the NE2001 model, there is significant evidence that the turbulent power spectrum deviates from the Kolmogorov value along several lines of sight, and this in turn alters the spectral dependence of the temporal smearing time.   For instance, L\"ohmer et al. (2001) show that $\tau \propto \nu^{3.44 \pm 0.13}$ along certain lines of sight at dispersion measures exceeding $\sim 10^3\,$pc\,cm$^{-3}$.  Thus, along certain lines of sight, the temporal smearing time will be smaller than predicted at frequencies below the fiducial scaling frequency (of 1\,GHz assumed by the NE2001 model), and larger than predicted above this frequency. In this case, the smearing time in a scattering model scaled to a fiducial frequency $\nu_0$ but evaluated at a frequency $\nu$ is expected to deviate from the Kolmogorov value by a factor $\sim (\nu/\nu_0)^{\sim 0.9}$. This difference can be significant for certain lines of sight for $\nu/\nu_0 \gg 1$ or $\nu/\nu_0 \ll 1$.  We note that all of the quantities explicitly calculated in the figures in this paper are evaluated at the NE2001 default frequency of 1\,GHz.

Other uncertainties, such as those in the scattering measure, also impact the predicted event rate along a given specific line of sight.  However, it should be remembered that while the model may fail along specific lines of sight, its use as a broad predictor of the event rate across the Galaxy is nonetheless valid as long as the number of ``anomolous'' scattering regions is small\footnote{If it is not, we note that in the code used to calculate event rates (see \S\ref{sec:conc}) and available online, provision is made to alter the power law index of the temporal smearing.}.

Temporal smearing has a large impact on the event detection rate on impulsive transients at low frequencies or in highly-scattered regions of the Galaxy.  In these regions the scattering can be sufficiently strong that it effectively creates a sensitivity horizon beyond which it is difficult to detect impulsive emission.  This is particularly apparent for lines of sight that intersect the Galactic Plane and the Galactic Center region. 

\begin{figure}[htb!]
\begin{center}
\epsfig{file=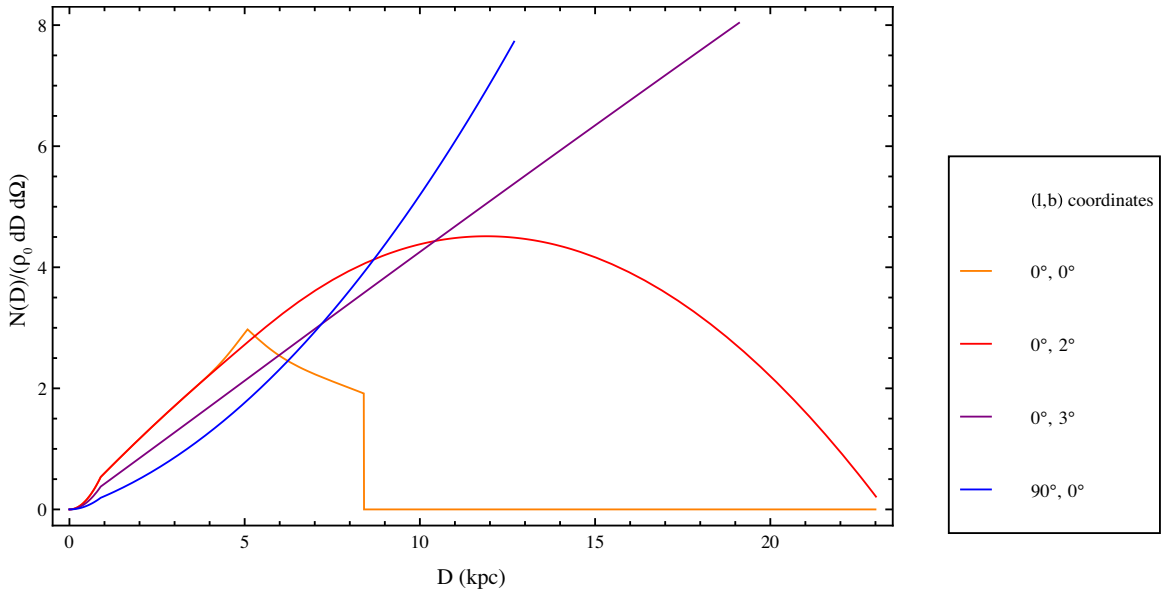,scale=0.7} \\
\caption{The normalized differential event rate $N(D)/\rho_0 d\Omega dD$ as a function of distance from the observer for various lines of sight through the Galaxy.  In the case $(l,b)=(0,0)$ the event rate drops to zero as the line of sight intersects the Galactic Center.  The other curves show how the event rate is affected progressively less as the line of sight moves away from the Galactic Center.  Each curve stops at the point at which it intersects the edge of the Galaxy.
The event rate is shown for parameters ${\cal L}_{\min} = 0.1$Jy\,kpc$^2$, ${\cal L}_{\max} = 100$Jy\,kpc$^{2}$, $\alpha = 1.5$, $R_{\max}=15\,$kpc, $R_c=8.5\,$kpc, $h=1.0\,$kpc.  Here $S_0 = 0.01\,$Jy, $\nu=1\,$GHz and the intrinsic duration of each transient is $\Delta t=5\,$ms.} 
\label{figGalacticCenter}
\end{center}
\end{figure}

Figure \ref{figGalacticCenter} demonstrates the horizon effect for objects that lie beyond the Galactic Center relative to other lines of sight through the Galaxy for the same set of parameters used in Figs.\ref{figGalacticEvents}.  The line of sight pointed directly at the Galactic Centre (GC) suffers the largest detection rate decrement due to scattering.  The detection rate of transient objects per unit distance declines rapidly beyond a distance of 5\,pc, and it is zero beyond the GC region, at $D\approx 8.2\,$kpc.  The detectability of transients exhibits a strong dependence on the details of the scattering: detection beyond the GC region is possible along a sight line aimed just 2 degrees above the GC, and a sight line a further degree higher suffers even less from the effects of scattering.

\begin{figure}[htb!]
\begin{center}
\epsfig{file=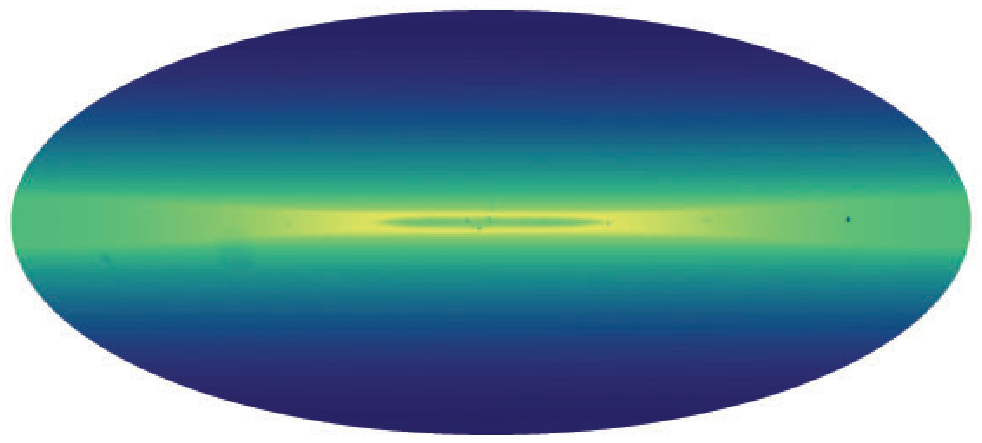,scale=1.6} \\
\epsfig{file=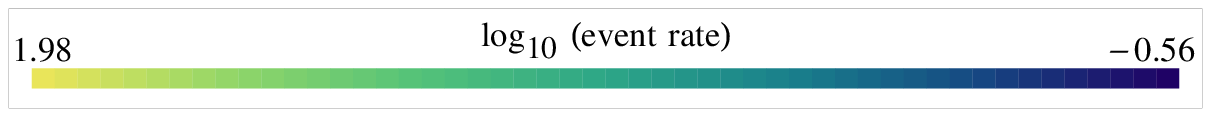,scale=0.7} \\
\caption{The normalized event rate in the Galaxy ${\cal R}/\rho_0 d\Omega$ for $L_{\min} = 0.1$Jy\,kpc$^2$, $L_{\max} = 100$Jy\,kpc$^{2}$, $\alpha = 1.5$, $R_{\max}=15\,$kpc, $R_c=8.5\,$kpc, $h=1.0\,$kpc.  The event rate is colour coded according to the log$_{10}$ of the event rate.  Here $S_0 = 0.01\,$Jy, $\nu=1\,$GHz and the intrinsic duration of each transient is $\Delta t=5\,$ms.} \label{figGalacticScatEvents}
\end{center}
\end{figure}

By integrating curves such as those seen in Fig.\,\ref{figGalacticCenter} over distance we derive the total rate of detectable transients for each line of sight through the Galaxy.   This is plotted in Fig.\,\ref{figGalacticScatEvents}, which vividly demonstrates the effect of temporal smearing on the event rate for certainly highly turbulent lines of sight through the Galaxy.

\section{Implications for Transients Surveys} \label{sec:discussion}

Surveys using an interferometric array can be broadly classified into modes that combine the array element outputs either incoherently or coherently, so that the survey searches for events using either the total power, or using interferometric visibilities.  We first consider below the relative merits of two most common ways of searching for transient events using incoherent combination of array outputs, before examining the circumstances in which a coherent search mode facilitates a superior event detection rate.  As each of these modes represents a different tradeoff between array sensitivity and access to FoV, we are primarily concerned here with the balance between survey depth and breadth that yields the optimal detection rate.  

There are several aspects of the detection of fast transients at radio wavelengths that distinguish it from transients surveys at other wavelengths.   Nemiroff (2003) considers the conditions for an optical transients survey in which is it better to tile a region of sky or stare at a single field.
However, for fast transients the duration of the event is small compared to telescope stare time, so there is no advantage to be gained by adopting one approach over the one.  Moreover, for homogeneously distributed extragalactic transients, one patch is as good as any other, while for Galactic transients, it is better to instantaneously cover the regions that yield the highest expected event rate.  

Another important difference is that in the optical regime there is no capability to distribute collecting area across multiple fields of view: there is only one telescope and it can only point in one direction at a time.  The larger flexibility afforded by a radio array allows greater scope to optimize the event detection rate.

\subsection{The competition between survey depth and breadth for the incoherent combination of telescope power} \label{secBreadthDepth}

We consider here the relative merits of the two forms of incoherent detection discussed earlier.  We consider the collimated survey mode in which all elements of the array point in the same direction and the total powers of the telescope are combined.  The survey FoV is that of the primary beam of an array element, $\Omega_t$, and the minimum detectable source flux density scales proportional to $S_0=N^{-1/2} {\cal S}$, where ${\cal S}$ is the flux density that would be detectable with a single array element.  In the opposite extreme, in the fly's eye mode all array elements are pointed in different directions, so it sacrifices survey depth in favor of breadth: the minimum detectable flux density is only $S_0={\cal S}$, but the FoV is now $N \Omega_t$. 

The tradeoff between survey depth and breadth depends on the functional dependence of the survey rate on survey FoV and sensitivity.  This dependence is ${\cal R} \propto \Omega S_0^{-3/2}$ for a sensitivity limited survey, which applies to homogeneously distributed extragalactic sources or a shallow survey in our Galaxy in which the effects of temporal smearing are negligible.  The collimated survey mode probes transients to a volume that is $N^{3/4}$ deeper than a single array element.  However, although in the fly's eye mode the survey probes a shallower volume, out to only the depth seen by a single element, it probes a volume of space $N$ times broader than a single element.  Thus we conclude that it is preferable to use array elements to cover as large a FoV as possible to maximize detection rate in a sensitivity-limited survey.

In a Galactic survey, the detection rate can exhibit a complicated dependence on survey sensitivity that alters the tradeoff between survey depth and breadth.  If the detection rate can be considered roughly uniform across the survey area (e.g. if the total FoV does not encompass too large a fraction of the sky), the detection rate takes the form ${\cal R} \propto \Omega S_0^{-3/2 + \delta}$, where $\delta$, in general, deviates from the value of zero that applies to a sensitivity-limited survey by virtue of interstellar scattering and the geometry of the Galaxy.  If $\delta < -1/2$ the collimated configuration becomes the optimal survey mode, while a value of $\delta > 0$ favors the fly's-eye mode even more strongly than for the sensitivity-limited survey considered above, with a relative advantage of $N^{1/4 + \delta/2}$.
   
\begin{figure}[htb!]
\begin{center}
\epsfig{file= 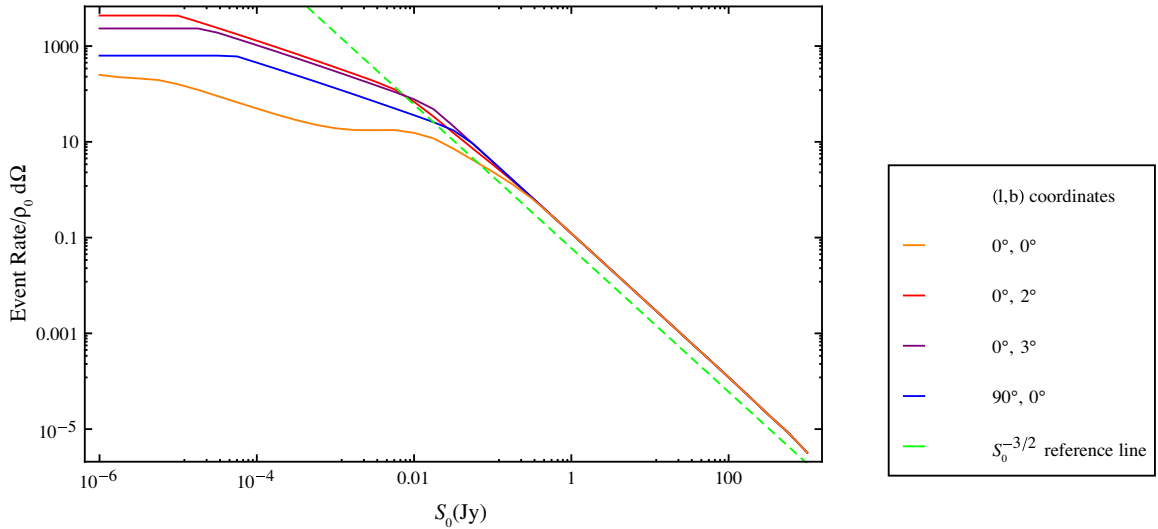,scale=0.7} \\
\caption{The normalized event rate ${\cal R}/\rho_0 d\Omega$ for various Galactic lines of sight  as a function of minimum detectable flux density, $S_0$, for ${\cal L}_{\min} = 0.1$Jy\,kpc$^2$, ${\cal L}_{\max} = 100$Jy\,kpc$^{2}$, $\alpha = 1.5$, $R_{\max}=15\,$kpc, $R_c=8.5\,$kpc, $h=1.0\,$kpc.  Here $\nu=1\,$GHz and the intrinsic duration of each transient is $\Delta t=5\,$ms.  None of these curves are steeper than $S_0^{-3/2}$.} \label{figGalacticRatevsSmin}
\end{center}
\end{figure}

The dependence of event rate on the limiting survey flux density is highly sensitive to the line of sight chosen. In general, one must plot ${\cal R}$ against $S_0$ in the range of interest to determine this for the given observing frequency and sight line, and the specific properties of the transient population.  Figure \ref{figGalacticRatevsSmin} shows the detection rate for four different lines of sight through the Galaxy for the same population parameters plotted in 
Fig.\,\ref{figGalacticCenter}.     It is apparent that $\delta$ exceeds zero for all lines of sight over the entire range of $S_0$, and it is as large as $\delta = 3/2$ at low values of $S_0$.    
 
Both scattering and the geometry of the Galaxy force $\delta$ to be positive always.  Scattering influences this dependence because, while greater sensitivity increases the proportion of events visible at large distance, these very objects are more susceptible to greater temporal smearing, which in turn decreases their detectability.  Thus a survey detects fewer objects at lower flux density than it otherwise would in the absence of scattering.  We conclude that scattering always moderates the dependence of event rate on limiting flux density, and $\delta > 0$.  This effect is particularly apparent in the behaviour of the $(l,b)=(0^\circ,0^\circ)$ curve in Fig.\,\ref{figGalacticRatevsSmin} in the range $10^{-6}\,{\rm Jy} \la S_0 \la 0.01\,$Jy; the curve shows a complex behaviour that depends on the nature of the scattering, but it is never steeper than $S_0^{-3/2}$.  

Effects related to the finite boundary of the Galaxy also force $\delta$ to be positive.  Once the survey sensitivity has increased to the point at which it detects objects at the boundary of the Galaxy, one begins to run out of sources, and further increases in survey sensitivity yield a lower increase in detections relative to the ${\cal R} \propto S_0^{-3/2}$ dependence associated with a sensitivity-limited survey.  At sufficiently low $S_0$ the survey ultimately finds all objects that would be detectable along that line of sight, at which point the event rate becomes insensitive to the survey sensitivity.  This effect is evident in the flattening of the three uppermost curves in the range $S_0 \sim 10^{-5}$\,Jy in Fig.\,\ref{figGalacticRatevsSmin}.  The mild break in slope observed over the range $10^{-5} \,{\rm Jy} \la  S_0 \la 0.1\,$Jy is also attributable to geometry: at $S_0 \sim 0.1\,$Jy the surveys already detect the most luminous objects at the edge of the Galaxy, and an increase in sensitivity yields no further detections of these events.  At progressively lower values of $S_0$ the survey runs out of further events at correspondingly lower luminosities.

\subsection{Strategy when the event rate varies strongly with Galactic coordinates}

The tradeoff between survey depth and breadth is altered if the detection rate varies appreciably across the FoV accessible by the entire array.  Figures \ref{figGalacticEvents}-\ref{figGalacticScatEvents} demonstrate that the event rate changes quickly as a function of position over certain regions of the sky.  We consider the circumstances under which it is favorable to concentrate collecting area on regions of high detection rate and thus probe deeper rather than broader.  Suppose that the survey area consists of two regions, with whose event rates are proportional to ${\cal E}_1\,$ster$^{-1}\,$Jy$^{3/2 - \delta/2}$ and ${\cal E}_2\,$ster$^{-1}\,$Jy$^{3/2 - \delta/2}$, with the ratio of these rates $\eta = {\cal E}_1/{\cal E}_2 \geq 1$.   The region with high event rate, region 1, subtends a solid angle $\Omega_1$ which is small enough that a fraction, $f < 1$, of the array elements suffice to cover this entire region in a fly's-eye mode.  

We would like to know how to distribute array elements to maximize the event detection rate.   The arguments in \S\ref{secBreadthDepth} reveal that the survey should at least tile the entirety of region 1 first , so the question becomes whether one should (i) place the remaining $(1-f)N$ elements across some of region 2 in order to survey more broadly, or (ii) use the remaining array elements to duplicate pointings in region 1 and thus increase sensitivity in the region with the higher event rate?  

The optimal detection strategy depends only on the values of $\eta$ and $f$.   Compare the event rates for the two survey strategies.  The fraction of array elements that are needed to cover region 1 once is $f = \Omega_1/N\Omega_t$, where $\Omega_t$ is the FoV of each array element.  In the collimated survey mode it is thus possible to tile this region $n=\lfloor f^{-1} \rfloor$ times (i.e. each individual pointing is observed by $n$ array elements).  We henceforth assume purely for the sake of simplicity that region 1 is tiled by an integral number of array element beams, and that $n$ is an integer divisor of $N$.  Then the detection rate for a survey that concentrates on region 1 alone can be written in the form,
\begin{eqnarray}
{\cal R}_{\rm col} = A {\cal E}_1 \Omega_1 \left( S_0/n^{1/2} \right)^{-3/2+\delta},  
\end{eqnarray}
where $A$ is a constant that depends on the luminosity distribution, interstellar scattering and the geometry of the Galaxy.  If, on the other hand, the remaining array elements are pointed at region 2, the detection rate is
\begin{eqnarray}
{\cal R}_{\rm fly's-eye} &=& A {\cal E}_1 \Omega_1 S_0^{-3/2+\delta} + A {\cal E}_2 (1-f) N \Omega_t S_0^{3/2+\delta} \nonumber \\
&=& A {\cal E}_1 \Omega_1 S_0^{-3/2+\delta} + A \frac{{\cal E}_1}{\eta} (1-f) \frac{\Omega_1}{f} S_0^{3/2+\delta}, 
\end{eqnarray}
where we have assumed that the value of $\delta$ is a constant over the range of flux densities $S_0 \, n^{-1/2} < S < S_0$.

Comparison of the two rates shows that it is preferable to concentrate all the collecting area on region 1 when the following condition is satisfied,
\begin{eqnarray}
n^{3/4 - \delta /2} > 1 + \frac{1-f}{\eta\,f}. \label{surveyIneq}
\end{eqnarray}
Using the definition $n=\lfloor f^{-1} \rfloor$ to rewrite the LHS of eq.\,(\ref{surveyIneq}), it is apparent that the conditions under which collimated pointings on region 1 alone are preferable depends on the balance between the ratio of event rates in the two regions against the size of the survey region. 
To illustrate the behaviour of the inequality, consider the case for $\delta = 0$.
If $f=0.5$ it is preferable to collect array elements in region 1 once the ratio $\eta$ exceeds 1.47 (i.e. our survey should consist of every pair of antennas pointing in the same direction).  If $f=0.33$ the collimated survey mode is preferable once $\eta > 1.57$ and array elements should be collected in groups of three, with a total of $N/3$ distinct pointings over region 1.  For $f=0.25$ the collimated survey mode is preferable for $\eta > 1.64$.  In summary, we see that as the size of region 1 decreases, a corresponding increase in rate disparity between the two survey regions is required for the collimated survey mode to remain the optimal detection method. 

The behaviour of this inequality corresponds to intuition in two obvious limiting cases.  
In the case $\eta =1$ we see the inequality in eq.\,(\ref{surveyIneq}) is never satisfied (except for unphysical range $f>1$), which reproduces the known result that it is always preferable to distribute array elements over both regions 1 and 2 when the event rate is identical in the two regions.  In the opposite extreme, if $\eta \gg 1$, the large disparity in event rates suggests the survey should always opt to concentrate all of its array elements in region 1 at the expense of region 2.


\subsection{Coherent, Visibility-based detection}

Coherent detection, in which all elements of the array are pointed at a common sky position and the interferometric visibilities are searched for transients, offers the highest sensitivity possible from the array, with $S_0 \propto N^{-1}$, but it does not usually allow access to the largest FoV.  The sensitivity comes at the price of extreme computational expense: one must search an extremely large number of synthesized beams in order to cover the entire FoV of each array element.   For a dish aperture diameter or station size $d$, the size of the synthesized beam is of order $\Omega_{\rm synth} = \pi (\lambda/d)^2$ which is usually much less than than the element FoV, $\Omega_t$.  Under many circumstances computational reality restricts the survey to a small number, $\xi$, of the synthesized beams that cover the entire element beam, so that $\xi \Omega_{\rm synth} \ll \Omega_t$.  

The ratios of the event rates in the coherent mode to other survey modes are,
\begin{eqnarray}
\frac{{\cal R}_{\rm coher}}{{\cal R}_{\rm fly's-eye}} 
= \xi \frac{\Omega_{\rm synth}}{\Omega_t} N^{1/2 - \delta},  \label{RcohervsFE}
\end{eqnarray}
and,
\begin{eqnarray}
\frac{{\cal R}_{\rm coher}}{{\cal R}_{\rm col}} 
= \xi \frac{\Omega_{\rm synth}}{\Omega_t} N^{3/4 - \delta/2}, \label{RcohervsCol} 
\end{eqnarray}
where again we assume that the value of $\delta$ is a constant over the range of sensitivities in question, $S_0 \, n^{-1} < S < S_0$; if not, the average value of $\delta$ over this range should be adopted.

The coherent detection mode constitutes the optimal survey strategy if $N$ is sufficiently large to overcome the disadvantage of a relatively small survey FoV.  This is particularly applicable to large $N$ arrays, such as the SKA, when $\delta \approx 0$ (i.e. when scattering is unimportant and the survey is sensitivity limited).  As such, this advantage always applies to surveys of extragalactic objects.  

However, an important caveat applies if the survey sensitivity is such that interstellar scattering effects are important or the survey probes events to the edge of the Galaxy, causing $\delta > 1/2$.  In this eventuality eq.\,(\ref{RcohervsFE}) shows that coherent detection mode becomes progressively {\it less} effective than the fly's-eye mode as $N$ increases.  The increase in sensitivity afforded by the coherent search mode is negated by the fact that event rate rises less steeply than $N$, and it is more effective to instead distribute the array elements into a fly's-eye configuration that increases the event rate $\propto N$.




\section{Conclusions} \label{sec:conc}

A summary of the main results of this paper is as follows:
\begin{itemize}
\item There is a critical difference between the Survey Figure of Merit for a telescope, which effectively measures survey speed, and the expected detection rate of fast transients.    This is fundamentally because one cannot trade integration time for sensitivity in a transients survey, so the weighting between sensitivity and FoV is different for the two metrics.
\item Surveys for extragalactic transients are ``sensitivity-limited'', and the distribution of distances of detected events depends only on survey sensitivity and the luminosity function of the events.  A luminosity function with a slope steeper than $-2$ preferentially detects events at low distances, whereas a distribution with a slope shallower than this preferentially detects events at the upper radius of the survey volume.  The detailed distance distribution can be used to optimize dedispersion engines used in the detection of fast transients in such surveys.
\item Transients surveys for Galactic populations are complicated by the fact that the survey volume has a finite extent.  At low sensitivity the survey is sensitivity-bound (as for the extragalactic case), but at high sensitivity the survey can become ``volume-bound'', in which the survey detects all events that exist out to the extent of the Galaxy.  At intermediate sensitivities, the survey is volume-bound for high luminosity events and sensitivity bound for low luminosity events.  The interplay between the shape of the Galaxy, luminosity function and survey sensitivity leads to a rich dependence of expected event rate with Galactic position.
\item The introduction of interstellar scattering hampers the detection of short-duration transients and, due to the highly inhomogeneous distribution of turbulent plasma within the Galaxy, further complicates the dependence of event rate on Galactic position.  Interstellar temporal broadening decrements the expected event rate to a large degree along heavily scattered lines of sight, particularly those in the Galactic plane and especially towards the Galactic Centre.  For some sight lines interstellar scattering acts as a barrier, at distances beyond which the detection of transients is either difficult or effectively impossible (e.g. within a few degrees of the Galactic Centre).  The effects of interstellar scattering must be integrated numerically in the context of a model of the Galactic electron distribution, and this leads to an extremely rich dependence of expected event rate on sky position.
\item In a survey with minimum detectable flux density $S_0$ and FoV $\Omega$ the detection rate scales as $\Omega \, S_0^{-3/2 + \delta}$ where $\delta =0$ for an extragalactic survey and any sensitivity-limited survey in which temporal broadening is unimportant.  In the Galaxy, both interstellar scattering and the finite extent of the Galaxy force the correction index $\delta$ into the range $0 \leq \delta \leq 3/2$.
\item For a survey in which expected event rate is constant over the FoV a fly's-eye survey detects a factor $N^{1/4}$ more events than a survey in which all $N$ elements of the array point at the same patch of sky and the array element outputs are combined incoherently (i.e. total powers are summed).
\item  For a survey in which there are large variations in the event rate as a function of sky position, one should conduct a collimated incoherent survey on the high event rate region instead of a fly's-eye survey over both high and low event rate regions when the approximate inequality $f^{3/4 - \delta /2} > 1 + \frac{1-f}{\eta\,f}$ holds.  The quantity $N f$ is the ratio of the solid angle covered by the high event rate region to the array element FoV (i.e. $\Omega_1/\Omega_t$), and $\eta>1$ is the ratio of the high to the low event rate per solid angle.
\item A survey in which the array outputs are combined coherently is the optimal detection method when 
$\xi \Omega_{\rm synth} N^{1/2-\delta} / \Omega_t > 1$, where $\xi \Omega_{\rm synth}$ is the FoV that can be processed by the survey and $\Omega_t$ is the array element FoV.  For high sensitivity Galactic surveys there may be circumstances in which scattering (or, less likely, the finite extent of the Galaxy) causes $\delta > 1/2$ and a coherent survey is {\it never} the optimal survey strategy, despite its advantage in sensitivity relative to incoherent survey modes.  The effects of temporal smearing become increasingly important at frequency, scaling as $\nu^{-4.4}$, so that a coherent survey can be sub-optimal at low frequency (e.g. 100\,MHz), even if the sensitivity is low.  
\end{itemize}

A copy of the {\tt Mathematica} code used to compute the event rates, including the effect of scattering, is available at {\tt https://safe.nrao.edu/vlba/vfastr/EventRatesCalculator.nb}.

\acknowledgments

The author is grateful to R.D.~Ekers, D.A.~Frail, P.~Hall and C.~Trott for their many and various comments and suggestions.  The Centre for All-sky Astrophysics is an Australian Research Council Centre of Excellence, funded by grant CE11E0090.



\clearpage

\end{document}